\documentclass[preprint,showpacs,preprintnumbers,amsmath,amssymb]{revtex4}

\usepackage{graphicx}
\usepackage{dcolumn}
\usepackage{bm}

\begin{document}

\preprint{DISTA FIN 02 2003}

\title{Uncoupled continuous-time random walks: solution and limiting behaviour
of the master equation}

\author{Enrico Scalas}
\affiliation{Department of Advanced Sciences and Technologies, East Piedmont University, Corso Borsalino 54, I-15100 Alessandria, Italy.}
\email{scalas@unipmn.it}
\homepage{http://www.fracalmo.org}

\author{Rudolf Gorenflo}
\affiliation{First Mathematical Institute, Free University of Berlin, Arnimallee 2-6, D-14195 Berlin, Germany}

\author{Francesco Mainardi}
\affiliation{Department of Physics, Bologna University and INFN Bologna, via Irnerio 46,
I-40126 Bologna, Italy}

\date{\today}

\pacs{05.40.-a, 89.65.Gh, 02.50.Cw, 05.60.-k, 47.55.Mh}

\begin{abstract}
A detailed study is presented for a large class of uncoupled continuous-time
random walks (CTRWs). The master equation is solved for the Mittag-Leffler 
survival probability. The properly scaled diffusive limit of the master equation
is taken and its relation with the fractional diffusion equation is discussed.
Finally, some common objections found in the literature are thoroughly reviewed. 
\end{abstract}

\maketitle

\section{Introduction}

The idea of combining a stochastic process for waiting times between two consecutive events and another stochastic process which associates a reward or a claim to each event dates back at least to the first half of the XXth Century \cite{lundberg03,cramer30}. The Cram\'er--Lundberg model for insurance risk is based on a point (or renewal) process \cite{cox67,cox79} ruling the random times at which random claims have to be paid by the company due to the occurence of accidents. Similar concepts have been used in renewal theory and in queueing theory, as well
\cite{cox67,cox79,khintchine60,gnedenko68,feller71,ross97}.

In the 1960s, Montroll and Weiss published a celebrated series of papers on random walks, where they applied the ideas developed by mathematicians working in probability theory to the physics of diffusion processes on lattices. In particular, they wrote a paper on continuous-time random walks (CTRWs) \cite{montroll65}, in which the waiting-time between two consecutive jumps of a diffusing particle is a real positive stochastic variable. 

The paper of Montroll and Weiss on CTRWs was the starting point for several developments on the physical theory of diffusion. In more recent times, CTRWs were applied back to economics and finance by Rudolf Hilfer \cite{hilfer84}, by the authors of the present paper \cite{scalas00,mainardi00,gorenflo01,raberto02} and, later, by Weiss and co-workers
\cite{masoliver03a,masoliver03b} and by Kutner {\it et al.} \cite{kutner03a,kutner03b}. However, here, the focus will be on anomalous relaxation properties of the waiting-time probability density and on the consequent relation between CTRWs and fractional diffusion.

Anomalous relaxation with power-law tails of the waiting-time density was investigated by means of Monte Carlo simulation by Montroll and Scher \cite{montroll73}. Shlesinger, Tunaley and other authors studied the asymptotic behaviour of CTRWs for large times \cite{shlesinger74,tunaley74,tunaley75,tunaley76,shlesinger82} (see also
\cite{weiss83}). Hilfer has recognized the important role played by Mittag-Leffler type functions in anomalous relaxation \cite{hilfer95a,hilfer03a}. Interesting contributions about the theory of CTRWs, including the problem of anomalous relaxation, can be found in refs. \cite{klafter87,klafter90,metzler94,kotulski95,klafter96,metzler99,barkai00,metzler00,uchaikin00,barkai01,sokolov01,barkai02,huillet02a,huillet02b}. The recent book by ben-Avraham and Havlin discusses in depth the possible applications of the formalism developed in the aforementioned papers \cite{ben-avraham00}. 

The asymptotic relation between properly scaled CTRWs with power-law waiting times and fractional diffusion processes has already been rigorously studied by Balakrishnan in 1985, dealing with anomalous diffusion in one dimension \cite{balakrishnan85}, four
years before the publication of the fundamental paper by Schneider and Wyss on the analytic theory of fractional diffusion and wave equations \cite{schneider89}. Later, many authors discussed this relation \cite{fogedby94,roman94,balescu94,hilfer95b,compte96,saichev97,hilfer00,meerschaert02,
scalas03}. As written above, the correspondence between CTRWs with Mittag-Leffler waiting time and the time-fractional diffusion equation has been lucidly worked out and explained in ref. \cite{hilfer95b} by Hilfer and Anton, who have shed light on the relevance of the Mittag-Leffler function, their specific aims, methods and interpretations being completely different from those of Balakrishnan. However, it must be recognized that already Balakrishnan in his formula (27) has found, as the natural choice for the waiting-time in CTRWs approximating fractional diffusion, the waiting-time density whose Laplace transform is (in the notation used in this paper) $1/(1+cs^{\beta})$, where $c$ is a positive constant. Implicitly, this is the Mittag-Leffler waiting-time described in Sec. III below. 
Meerschaert {\it et al.} have developed a method to derive the equations for CTRWs in the diffusive limit \cite{meerschaert02}. In their paper, they discuss both the coupled and uncoupled case. 

The present paper is devoted to a detailed discussion of the uncoupled case and it is organized as follows. In Sec. II, the basic quantities are introduced and a summary of the theory is given. Sec. III is devoted to the solution of the master equation in the uncoupled case. General formulae are presented and specialized to the case of the Mittag-Leffler waiting-time survival probability, in which an exact solution is available in terms of a fractionally generalized compound-Poisson process. In this section, a fractional relaxation equation satisfied by the Mittag-Leffler function is discussed. In Sec. IV, the proper scaling leading to the fractional diffusion equation is presented. The main result of this section is that the solution of the CTRW master equation weakly converges to the solution of a Cauchy problem for the fractional diffusion equation. This result is a version of the Central Limit Theorem and the steps for a rigorous proof are sketched. Finally, in Sec. V, the reader can find a discussion of some objections which are usually raised when dealing with fractional diffusion. Unnecessary mathematical difficulties have been avoided throughout the paper.

\section{Basic definitions}

As mentioned in the introduction, CTRWs are essentially point processes with reward. The point process is characterized by  a sequence of independent identically distributed (i.i.d.) positive random variables $\tau_i$, which can be interpreted as waiting times between two consecutive events:
\begin{equation}
\label{timewalk}
t_n = t_0 + \sum_{i=1}^{n} \tau_i; \; \; t_n - t_{n-1} = \tau_n; \; \; n=1, 2, 3, \ldots;
\; \; t_0 = 0. \end{equation} 
The rewards are (i.i.d.) not necessarily positive random variables: $\xi_i$. In the usual physical intepretation, the $\xi_i$s represent the jumps of a diffusing particle (the walker), and they can be $n$-dimensional vectors. In this paper, only the 1-dimensional case is studied, but the extension of many results to the $n$-dimensional case is straightforward. The position $x$ of the walker at time $t$ is (with $N(t) = \max \{ n:\;t_{n} \leq t \}$ and $x(0)=0$):
\begin{equation}
\label{jumpwalk}
x(t) = \sum_{i=1}^{N(t)} \xi_i.
\end{equation}
CTRWs are rather good and general phenomenological models for diffusion, including anomalous diffusion, provided that the time of residence of the walker is much greater than the time it takes to make a jump. In fact, in the formalism, jumps are instantaneous.

In general, the jumps and the waiting times are not independent from each other. Then, the random walk can be described by the joint probability density $\varphi (\xi, \tau)$ of jumps and waiting times; $\varphi (\xi, \tau) \, d \xi \, d\tau$ is the probability of a jump to be in the interval $(\xi, \xi+ \, d\xi)$ and of a waiting time to be in the interval $(\tau, \tau+ \, d\tau)$. The following integral equation gives the probability density, $p(x,t)$, for the walker being in position $x$ at time $t$, conditioned by the fact that it was in position $x=0$ at time $t=0$:
\begin{equation}
\label{masterequation}
p(x,t) =  \delta (x)\, \Psi(t) +
   \int_0^t \, 
 \int_{-\infty}^{+\infty}  \varphi(x-x',t-t')\, p(x',t')\, dt'\,dx',
\end{equation}
where $\Psi(\tau)$ is the so-called survival function. $\Psi(\tau)$ is related to the marginal waiting-time probability density $\psi(\tau)$. The two marginal densities $\psi(\tau)$ and $\lambda(\xi)$ are:
\begin{eqnarray}
\label{marginal}
\psi(\tau) & = & \int_{-\infty}^{+\infty} \varphi(\xi, \tau) \, d \xi \nonumber \\
\lambda(\xi) & = & \int_{0}^{\infty} \varphi(\xi, \tau) \, d \tau,
\end{eqnarray}
and the survival function $\Psi(\tau)$ is:
\begin{equation}
\label{survival}
\Psi(\tau) = 1 - \int_{0}^{\tau} \psi (\tau') \, d \tau' = \int_{\tau}^{\infty} \psi (\tau') \, d \tau'.
\end{equation}

The integral equation, eq. (\ref{masterequation}), can be solved in the Laplace-Fourier domain. The Laplace transform, $\widetilde{g}(s)$ of a (generalized) function $g(t)$ is defined as:
\begin{equation}
\label{laplacetransform}
\widetilde{g}(s) = 
\int_{0}^{+ \infty} dt \; 
\hbox{e}^{ -st} \, g(t)\,,
\end{equation}
whereas the Fourier transform of a (generalized) function $f(x)$ is defined as:
\begin{equation}
\label{fouriertransform}
\widehat {f}(\kappa) = 
\int_{- \infty}^{+ \infty} dx \, 
\hbox{e}^{i \kappa x} \, f(x)\,.
\end{equation}
A generalized function is a distribution (like Dirac's $\delta$) in the sense of S. L. Sobolev and L. Schwartz \cite{gelfand58}.

One gets:
\begin{equation}
\label{gensol}
\widetilde{\widehat p}(\kappa, s) = \widetilde \Psi(s)\, \frac{1}{1-\widetilde{\widehat \varphi}(\kappa, s)},
\end{equation}
or, in terms of the density $\psi(\tau)$:
\begin{equation}
\label{vargensol}
\widetilde{\widehat p}(\kappa, s) = \frac{1 -\widetilde \psi(s)}{s}\, \frac{1}{1-\widetilde{\widehat \varphi}(\kappa, s)},
\end{equation}
as, from eq. (\ref{survival}), one has:
\begin{equation}
\label{survivallt}
\Psi(s) = \frac{1 -\widetilde \psi(s)}{s}.
\end{equation} 
In order to obtain $p(x,t)$, it is then necessary to invert its Laplace-Fourier transform $\widetilde{\widehat p}(\kappa, s)$. Analytic solutions are quite important, as they provide a benchmark for testing numerical inversion methods. In the next section, an explicit analytic solution for a class of continuous-time random walks with anomalous relaxation behaviour will be presented. It will be necessary to restrict oneself to the uncoupled case, in which jumps and waiting-times are not correlated. 

\section{Solution of the master equation}

In this section, the solution of eq. (\ref{masterequation}) will be discussed in the uncoupled case. First of all, a general formula will be derived for $p(x,t)$, then it will be specialized to two cases: the well-known case of an exponential survival function and the case where the survival function is a Mittag-Leffler function. The connections and the analogies between these two cases will be presented. A new solution will be obtained in terms of a fractionally-generalized compound Poisson process.

As anticipated above, the study will be restricted to uncoupled continuous-time random walks. This means that jump sizes do not depend on waiting times and the joint probability density for jumps and waiting times can be factorized in terms of the two marginal densities:
\begin{equation}
\label{factorization}
\varphi(\xi, \tau) = \lambda(\xi) \psi (\tau)
\end{equation}
with the normalization conditions $\int d \xi \lambda (\xi) = 1$ and $\int d \tau \psi(\tau)=1$.

In this case the integral master equation for $p(x,t)$ becomes:
\begin{equation}
\label{uncoupledreal}
p(x,t) =  \delta (x)\, \Psi(t) +
   \int_0^t   \psi(t-t') \, \left[
 \int_{-\infty}^{+\infty}  \lambda(x-x')\, p(x',t')\, dx'\right]\,dt'
\end{equation}
This equation has a well known general explicit solution in terms of $P(n,t)$, the probability of $n$ jumps occurring up to time $t$, and of the $n$-fold convolution of the jump density, $\lambda_n (x)$: 
\begin{equation}
\label{jumpconv}
\lambda_n (x) = \int_{-\infty}^{+\infty} \, \int_{-\infty}^{+\infty} \ldots \int_{-\infty}^{+\infty} \,d \xi_{n-1} d \xi_{n-2} \ldots d \xi_1 \lambda(x - \xi_{n-1}) \lambda (\xi_{n-1}-\xi_{n-2}) \ldots \lambda(\xi_1).
\end{equation}
Indeed, $P(n,t)$ is given by:
\begin{equation}
\label{Poisson1}
P(n,t) = \int_{0}^{t} \psi_n (t - \tau) \Psi(\tau) \, d \tau
\end{equation}
where $\psi_n (\tau)$ is the $n$-fold convolution of the waiting-time density: 
\begin{equation}
\label{timeconv}
\psi_n (\tau) = \int_{0}^{\tau} \int_{0}^{\tau_{n-1}} \ldots \int_{0}^{\tau_1}
\, d \tau_{n-1} d \tau_{n-2} \ldots d \tau_1 \psi(t-\tau_{n-1}) \psi(\tau_{n-1} - \tau_{n-2}) \ldots \psi(\tau_1).   
\end{equation}
The $n$-fold convolutions defined above are probability density functions for the sum of $n$ variables.

The Laplace transform of $P(n,t)$, $\widetilde P(n,s)$, reads:
\begin{equation}
\label{Poisson1lt}
\widetilde P(n,s) = [ \widetilde \psi (s) ]^n \widetilde \Psi(s)
\end{equation}
By taking the Fourier-Laplace transform of eq. (\ref{uncoupledreal}), one gets:
\begin{equation}
\label{montroll}
\widetilde{\widehat p}(\kappa, s) = \widetilde \Psi(s)\,
\frac{1}{1- \widetilde \psi(s) \widehat \lambda(\kappa)}\,.
\end{equation}
But, recalling that $|\lambda(\kappa)| < 1$ and $|\psi(s)| < 1$, if $\kappa \not= 0$ and 
$s \not= 0$, eq. (\ref{montroll}) becomes:
\begin{equation}
\label{series1}
\widetilde{\widehat p}(\kappa, s) = \widetilde \Psi(s)\, \sum_{n=0}^{\infty} 
[\widetilde \psi(s) \widehat \lambda(\kappa)]^n \,;
\end{equation}
this gives, inverting the Fourier and the Laplace transforms and taking into account eqs. (\ref{jumpconv}) and (\ref{Poisson1}):
\begin{equation}
\label{series2}
p(x,t) = \sum_{n=0}^{\infty} P(n,t) \lambda_n (x)
\end{equation}
Eq. (\ref{series2}) can also be used as the starting point to derive eq. (\ref{uncoupledreal}) via the transforms of Fourier and Laplace, as it describes a jump process subordinated to a renewal process.

A remarkable analytic solution is available when the waiting-time probability density function has the following exponential form:
\begin{equation}
\label{exponential1}
\psi(\tau) = \mu \hbox{e}^{-\mu \tau}.
\end{equation}
Then, the survival probability is $\Psi (\tau) = \hbox{e}^{-\mu \tau}$ and the probability of $n$ jumps occurring up to time $t$ is given by the Poisson distribution:
\begin{equation}
\label{Poissondistribution}
P(n,t) = \frac{(\mu t)^n}{n!} \, \hbox{e}^{-\mu t}.
\end{equation}
In this case, equation (\ref{series2}) becomes:
\begin{equation}
\label{Lundbergsolution}
p(x,t) = \sum_{n=0}^{\infty} \frac{(\mu t)^n}{n!} \, \hbox{e}^{-\mu t} \lambda_n (x).
\end{equation}
When $\lambda(x)$ is the jump density for a positive random variable, eq. (\ref{Lundbergsolution}) is the starting point of the Cram\'er--Lundberg model for insurance risk \cite{lundberg03,cramer30}.
It is worth noting that the survival probability $\Psi(\tau)$ satisfies the following relaxation ordinary differential equation:
\begin{equation}
\label{normrelax}
\frac{d}{d \tau} \Psi(\tau) = - \mu \Psi(\tau), \,\,\, \tau>0; \,\,\, \Psi(0^+) =1.
\end{equation}

The simplest fractional generalization of eq. (\ref{normrelax}) giving rise to anomalous relaxation and power-law tails in the waiting-time probability density can be written as follows, by appropriately choosing the time scale:
\begin{equation}
\label{anrelax}
\frac{d^\beta}{d \tau^\beta} \Psi(\tau) = - \Psi(\tau), \,\,\, \tau>0, \,\,\, 0 < \beta \leq 1; \,\,\, \Psi(0^+) =1,
\end{equation}
where the operator $d^\beta / dt^\beta$ is the Caputo fractional derivative, related to the Riemann--Liouville fractional derivative. For a sufficiently well-behaved function $f(t)$, the Caputo derivative is defined by the following equation, for $0<\beta<1$:
\begin{equation}
\label{Capder}
\frac{d^\beta}{dt^\beta} f(t) = \frac{1}{\Gamma(1-\beta)} \, \frac{d}{dt} \int_{0}^{t} \frac{f(\tau)}{(t-\tau)^\beta}\,d\tau - \frac{t^{-\beta}}{\Gamma(1-\beta)}f(0^{+}),
\end{equation}
and reduces to the ordinary first derivative for $\beta = 1$. The Laplace transform of the Caputo derivative of a function $f(t)$ is:
\begin{equation}
\label{Capderlt}
{\cal{L}} \left( \frac{d^\beta}{dt^\beta} f(t);\,s \right) = s^{\beta} \widetilde f (s) - s^{\beta - 1} f(0^+).
\end{equation}
If eq. (\ref{Capderlt}) is applied to the Cauchy problem of eq. (\ref{anrelax}), one gets:
\begin{equation}
\label{anrelaxlt}
\widetilde \Psi (s) = \frac{s^{\beta - 1}}{1+s^\beta}.
\end{equation}
Eq. (\ref{anrelaxlt}) can be inverted, giving the solution of
eq. (\ref{anrelax}) in terms of the Mittag-Leffler function of parameter $\beta$ \cite{mainardi96,gorenflo97}:
\begin{equation}
\label{ML}
\Psi(\tau) = E_{\beta} (-\tau^{\beta}),
\end{equation}
defined by the following power series in the complex plane:
\begin{equation}
\label{MLseries}
E_{\beta} (z) := \sum_{n=0}^{\infty} \frac{z^n}{\Gamma(\beta n +1)}.
\end{equation}
For small $\tau$, the Mittag-Leffler survival function has the same behaviour as a stretched exponential:
\begin{equation}
\label{stretched}
\Psi(\tau) = E_{\beta} (-\tau^{\beta}) \simeq 1 - \frac{\tau^\beta}{\Gamma(\beta +1)} \simeq \exp \{-\tau^\beta / \Gamma(\beta +1) \}, \; 0 \leq \tau \ll 1,
\end{equation}
whereas for large $\tau$, it has the asymptotic representation:
\begin{equation}
\label{powerlaw}
\Psi(\tau) \sim \frac{\sin (\beta \pi)}{\pi} \, \frac{\Gamma(\beta)}{\tau^\beta}, \; 0<\beta<1,
\; \tau \to \infty.
\end{equation}
Accordingly, for small $\tau$, the probability density function of waiting times $\psi(\tau) = - d \Psi(\tau) / d \tau$ behaves as:
\begin{equation}
\label{stretchedd}
\psi(\tau) = - \frac{d}{d \tau} E_{\beta} (-\tau^{\beta}) \simeq \frac{\tau^{-(1-\beta)}}{\Gamma(\beta+1)} \exp \{-\tau^\beta / \Gamma(\beta +1) \}, \; 0 \leq \tau \ll 1,
\end{equation}
and the asymptotic representation is:
\begin{equation}
\label{powerlawd}
\psi(\tau) \sim \frac{\sin (\beta \pi)}{\pi} \, \frac{\Gamma(\beta+1)}{\tau^{\beta+1}}, \; 0<\beta<1, \; \tau \to \infty.
\end{equation}
Before going on, it is now time to review the results obtained so far. The solution of equation (\ref{anrelax}) is a survival probability function $\Psi(\tau)$ with power-law decay $\tau^{-\beta}$ if $0<\beta<1$ and $\tau \to \infty$. The decay exponent of the corresponding probability density function $\psi(\tau)$ is $-(\beta+1)$, with values in the interval $(1,2)$. This ensures that the normalization condition for $\psi(\tau)$ can be satisfied. However, already the first moment of $\psi(\tau)$ is infinite. It is worth stressing that the case $\beta=1$ does not correspond to a $\tau^{-1}$--decay of the survival probability, but to the exponential relaxation described by eq. (\ref{normrelax}).

The Laplace transform of $\psi(\tau)$ is given by (see eq. (\ref{survivallt})):
\begin{equation}
\label{dl}
\widetilde \psi(s) = 1 - s \widetilde \Psi(s) = \frac{1}{1+s^\beta}.
\end{equation}
Therefore, recalling eq. (\ref{Poisson1lt}) and eq. (\ref{anrelaxlt}), one can obtain the Laplace transform of $P(n,t)$:
\begin{equation}
\label{Genpoissonlt}
\widetilde P(n,s) = \frac{1}{(1+s^\beta)^n} \frac{s^{\beta - 1}}{1+s^\beta}.
\end{equation}
This can be analytically inverted as (see eq. (1.80) in ref. \cite{podlubny99}):
\begin{equation}
\label{podlubny}
{\cal{L}} [t^{\beta n} E_{\beta}^{(n)}(-t^\beta); s] = \frac{n!s^{\beta - 1}}{(1+s^{\beta})^{n+1}},
\end{equation}
where:
$$E_{\beta}^{(n)}(z) := \frac{d^n}{dz^n} E_{\beta} (z).$$
Eq. (\ref{podlubny}) yields an explicit analytic expression for $P(n,t)$:
\begin{equation}
\label{Genpoisson}
P(n,t) = \frac{t^{\beta n}}{n!} E_{\beta}^{(n)} (-t^\beta)
\end{equation}
Eq. (\ref{Genpoisson}) generalizes the Poisson distribution (\ref{Poissondistribution}) for the anomalous relaxation case under study ($0 < \beta < 1$). It reduces to the Poisson distribution in the case $\beta = 1$, in which the Mittag-Leffler function coincides with the exponential function. As an immediate consequence of this result and of eq. (\ref{series2}), one also gets the analytic solution of the master equation (\ref{uncoupledreal}) for a continuous-time random walk characterized by the survival function of eq. (\ref{ML}):
\begin{equation}
\label{exact}
p(x,t) = \sum_{n=0}^{\infty} \frac{t^{\beta n}}{n!} E_{\beta}^{(n)} (-t^\beta) \lambda_n(x).
\end{equation}
As a consistency check, one can show that
\begin{equation}
\label{normalization}
\int_{-\infty}^{+\infty} p(x,t) \, dx = 1, \; \forall t.
\end{equation}
This is equivalent to the requirement that the Fourier transform computed in $\kappa = 0$ satisfies $\widehat p(0,t) = 1, \; \forall t$. But $\widehat p(0,t)$ is given by:
\begin{equation}
\label{FT}
\widehat p(0,t) = \sum_{n=0}^{\infty} \frac{t^{\beta n}}{n!} E_{\beta}^{(n)} (-t^\beta)
\end{equation}
and recalling that for any sufficiently well-behaved function, $f$:
$$
f(a+\delta) = \sum_{n=0}^{\infty} \frac{f^{(n)}(a)}{n!} \delta^n
$$
identifying $a = -t^\beta$ and $\delta = + t^\beta$, one has the following chain of equalities:
\begin{equation}
\label{proof}
\widehat p(0,t) = \sum_{n=0}^{\infty} \frac{t^{\beta n}}{n!} E_{\beta}^{(n)} (-t^\beta) =
E_\beta ((-t^\beta) + t^\beta) = E_\beta (0) = 1.
\end{equation}
It is now interesting to investigate the behaviour of the exact solution given by eq. (\ref{exact}) in the so-called diffusive or hydrodynamic limit. This limit is obtained by making smaller all waiting times by a positive factor $r$, and all jumps by a positive factor $h$ and then letting $r$ and $h$ vanish in an appropriate way. This will be the subject of the next section.

\section{The diffusive limit}

In this section, for the first time, a collection of results by the authors of this paper is made available in a complete way; mathematical subtleties have been recalled wherever necessary. Partial results were discussed in refs. \cite{gorenflo01,scalas03,gorenflo02}. Here, the focus is on the well-scaled transition to the diffusive limit based on sound limit theorems of probability theory. The following derivation should help the reader in judging whether, in the problem he/she is dealing with, the connection between CTRWs and fractional diffusion is relevant.

As mentioned above, in order to discuss the diffusive limit, the waiting times are multiplied by a positive factor $r$ so that one gets:
\begin{equation}
\label{scalingtime}
t_n (r) = r \tau_1 + r \tau_2 + \ldots + r \tau_n.
\end{equation}
Analogously, the jumps are multiplied by a positive factor $h$. Letting $x_0 (h) = 0$, one has:
\begin{equation}
\label{scalingjumps}
x_n (h) = h \xi_1 + h \xi_2 + \ldots + h \xi_n.
\end{equation}
The probability density function $\psi_r (\tau)$ of the scaled waiting times is related to $\psi(\tau)$ in the following way:
\begin{equation}
\label{scaledpdft}
\psi_r (\tau) = \frac{\psi(\tau / r)}{r}, \; \tau > 0,
\end{equation}
The scaled-jump probability density function $\lambda_h(\xi)$ is given by:
\begin{equation}
\label{scaledpdfj}
\lambda_h (\xi) = \frac{\lambda (\xi / h)}{h}
\end{equation}
The Laplace transform of $\psi_r (\tau)$ and the Fourier transform of $\lambda_h(\xi)$ are, respectively:
\begin{equation}
\label{scaledtransfs}
\widetilde \psi_r (s) = \widetilde \psi (r s), \; \; \widehat \lambda_h (\kappa) = \widehat \lambda( h \kappa).
\end{equation}
In the Fourier-Laplace domain, the rescaled solution of the master equation reads:
\begin{equation}
\label{scaledsolution}
\widetilde{\widehat{p_{r,h}}} (\kappa, s) = \frac{1-\widetilde \psi_r (s)}{s}
\frac{1}{1 - \widetilde \psi_r (s) \widehat \lambda_h (\kappa)}.
\end{equation}
Eq. (\ref{scaledsolution}) will be the starting point for investigating the diffusive limit of the solution presented in eq. (\ref{exact}). The results discussed above, from eq. (\ref{scalingtime}) to eq. (\ref{scaledsolution}), are rather general. It is now possible to specialize them to the class of waiting-time densities discussed in Sec. III and to a large class of jump densities.

For $0< \beta <1$, eq. (\ref{powerlawd}) gives the asymptotic representation of the waiting-time density. For such a behaviour, one has, for each fixed $s>0$ that:
\begin{equation}
\label{scalings}
\widetilde \psi_r (s) = \widetilde \psi(r s) = 1 - c_1 (r s)^{\beta} + o (r^\beta), \; r \to 0.
\end{equation}
In the case under study, it turns out that $c_1 = 1$.
Remarkably, this result holds also for $\beta = 1$.
An important class of symmetric jump densities ($\lambda(-\xi) = \lambda(\xi)$) is characterized by the following behaviour, for $b > 0$ and some parameter $\alpha \in (0,2)$:
\begin{equation}
\label{powerlawbis}
\lambda(x) = [b + \epsilon(|x|)] |x|^{-(\alpha+1)},
\end{equation}
with $\epsilon(|x|) \to 0$ as $|x| \to \infty$. For these densities, exhibiting a power-law decay at infinity, the following asymptotic relation holds:
\begin{equation}
\label{scalingj}
\widehat \lambda_h (\kappa) = \widehat \lambda (h \kappa) = 1 - c_2 (h |\kappa|)^\alpha + o(h^\alpha), \; h \to 0,
\end{equation}
where the constant $c_2$ is given by:
\begin{equation}
\label{c2}
c_2 = \frac{b \pi}{\Gamma(\alpha+1) \sin (\alpha \pi / 2)}
\end{equation}
Eq. (\ref{scalingj}) is valid also for symmetric densities with finite second moment $\sigma$. In that case, one has $\alpha = 2$ and $c_2 = \sigma^2 /2$.
Both the results in eq. (\ref{scalings}) and in eq. (\ref{scalingj}) are less trivial than they seem. Indeed, in order to prove eq. (\ref{scalings}), it is necessary to use a corollary on Laplace transforms discussed in the classical book by Widder (see ref. \cite{widder46}, p. 182), whereas the proof of eq. (\ref{scalingj}) is tedious but can be distilled from chapter 8 of \cite{bingham87}. A simpler but weaker proof can be found in ref. \cite{gorenflo02}.

By using the asymptotics in eq. (\ref{scalings}) and in eq. (\ref{scalingj}) and replacing in eq. (\ref{scaledsolution}), it follows that:
\begin{equation}
\label{asympt1}
\widetilde{\widehat{p_{r,h}}} (\kappa, s) \sim \frac{c_1 r^\beta s^{\beta-1}}{c_1 r^\beta s^\beta + c_2 h^\alpha |\kappa|^\alpha}, \; \; r,h \to 0.
\end{equation}
Now, the following scaling relation can be imposed:
\begin{equation}
\label{einstein}
c_1 r^\beta = c_2 h^\alpha,
\end{equation}
yielding, for $r,h \to 0$:
\begin{equation}
\label{difflimitfl}
\widetilde{\widehat{p_{r,h}}} (\kappa, s) \to \frac{s^{\beta-1}}{s^{\beta}+|\kappa|^\alpha}.
\end{equation}
This limit coincides with the Laplace-Fourier transform of the Green function (or fundamental solution) for the following fractional diffusion Cauchy problem:
\begin{eqnarray}
\label{fractional}
\frac{\partial^\beta}{\partial t^\beta} u(x,t)  & = & \frac{\partial^\alpha}{\partial |x|^\alpha} u(x,t), \; \; 0 < \alpha \leq 2, \; \; 0 < \beta \leq 1, \nonumber \\
u(x,0^+) & = & \delta (x), \; \; x \in (-\infty, +\infty), \; \; t > 0,
\end{eqnarray}
where $\partial^\beta / \partial t^\beta$ is the Caputo derivative defined in eq. (\ref{Capder}), and $\partial^\alpha / \partial |x|^\alpha$ is the Riesz derivative: a pseudo-differential operator with symbol $-|\kappa|^\alpha$. Recalling eq. (\ref{Capderlt}), the Laplace-Fourier transform of $u(x,t)$ reads:
\begin{equation}
\label{greenfl}
\widetilde{\widehat{u}} (\kappa, s) = \frac{s^{\beta-1}}{s^{\beta}+|\kappa|^\alpha},
\end{equation}
and therefore, as anticipated, one has, for $r,h \to 0$ under the scaling relation
eq. (\ref{einstein}):
\begin{equation}
\label{difflimfrac}
\widetilde{\widehat{p_{r,h}}} (\kappa, s) \to \widetilde{\widehat{u}} (\kappa, s).
\end{equation}
In this passage to the limit, $\widetilde{\widehat{p_{r,h}}} (\kappa, s)$ and $\widetilde{\widehat{u}} (\kappa, s)$ are asymptotically equivalent in the Laplace-Fourier domain. Then, the asymptotic equivalence in the space-time domain between the master equation eq. (\ref{uncoupledreal}) and the fractional diffusion equation, eq. (\ref{fractional}) is ensured by the continuity theorem for sequences of characteristic
functions, after the application of the analogous theorem for sequences of Laplace transforms \cite{feller71}. Therefore, there is convergence in law or weak convergence for the corresponding probability distributions and densities. Here, weak convergence means that the Laplace transform and/or Fourier transform (characteristic function)
of the probability density function are pointwise convergent
(see for details ref. \cite{feller71}).
In other words, under the appropriate scaling, defined by eq. (\ref{einstein}), and in the limit $r,h \to 0$, the solution given in eq. (\ref{exact}) weakly converges to the Green function of the fractional diffusion equation eq. (\ref{fractional}):
\begin{equation}
\label{greenfunction}
u(x,t) = \frac{1}{t^{\beta / \alpha}} W_{\alpha, \beta} \left( \frac{x}{t^{\beta/\alpha}} \right),
\end{equation}
where $W_{\alpha, \beta} (u)$ is given by:
\begin{equation}
\label{scalingfunction}
W_{\alpha, \beta} (u) = \frac{1}{2 \pi} \int_{-\infty}^{+\infty} \; d \kappa \; \hbox{e}^{-i \kappa u} E_\beta (-|\kappa|^\alpha),
\end{equation}
the inverse Fourier transform of a Mittag-Leffler function \cite{metzler00,scalas03,mainardi01,zaslavsky02}.
In the case $\beta =1$ and $\alpha = 2$, the fractional diffusion equation reduces to the ordinary diffusion equation and the function $W_{2,1} (u)$ becomes the Gaussian probability density function evolving in time with a variance $\sigma^2 = 2t$. In the general case ($0 < \beta < 1$ and $0 < \alpha <2$), the function $W_{\alpha, \beta} (u)$ is still a probability density evolving in time, and it belongs to the class of Fox $H$-type functions that can be expressed in terms of a Mellin-Barnes integral as shown in details in ref. \cite{mainardi01}.

The scaling equation, eq. (\ref{einstein}), can be written in the following form, where $C$ is a constant:
\begin{equation}
\label{properscaling}
h = C r^{\beta/\alpha}.
\end{equation}
If $\beta = 1$ and $\alpha = 2$, one recognizes the scaling typical of Brownian motion (or the Wiener process). Indeed, this is the limiting stochastic process for the uncoupled continuous-time random walks with exponential waiting-time density and the class of jump densities with finite second moment. In all the other cases considered in this paper: $\beta \in (0,1)$ and $\alpha \in (0,2)$, the limiting process has a probability density function given by $u(x,t)$ in eq. (\ref{greenfunction}).

\section{Discussion and conclusions}

In this paper, the connection between a class of CTRWs with Mittag-Leffler survival probability and the fractional diffusion equation has been discussed. In Sec. III, an explicit solution of the master equation has been derived for long-tail processes with Mittag-Leffler survival function. As shown in Sec. IV, it turns out that, for this class, the solution of the CTRW master equation weakly converges to the solution of a Cauchy problem for the fractional diffusion equation. Although such weak convergence also occurs for the waiting-time densities whose Laplace transforms have an $s \to 0$ asymptotics $1 - c_1 s^{\beta} + o(s^{\beta})$ (see refs. \cite{balakrishnan85,gorenflo02}), the Mittag-Leffler waiting-time law deserves special attention as, without passage to the diffusion limit, it leads to a time-fractional master equation, just by insertion into the CTRW integral equation. This fact was discovered and made explicit for the first time in 1995 by Hilfer and Anton \cite{hilfer95b}. Therefore, this special type of waiting-time law (with its particular properties of being singular at zero, completely monotonic and long-tailed) may be best suited for approximate CTRW simulation of fractional diffusion. It must be stressed that both the results of Sec. III and IV are based on sound and original mathematical considerations. 

It is important to remark that the presence of the time Caputo fractional derivative (or equivalently of the Riemann-Liouville derivative) and of the space Riesz derivative in eq. (\ref{fractional}) is a {\it natural} consequence of the well scaled diffusion limit discussed in Sec. IV. This should be already clear from previous work on the relation between CTRWs and fractional diffusion (see, in particular ref. \cite{hilfer95b}). However, it is still often argued that there is an arbitrariness in the choice of the fractional operator in eq. (\ref{fractional}). If one uses different fractional operators, the physical meaning, if any, of these versions of eq. (\ref{fractional}) will be different. 

Another point has been raised on the physical meaning of eq. (\ref{fractional}). In particular, some authors consider the space fractional derivative unphysical due to its non-locality. An answer to this objection is that it is always possible to use an equation as a phenomenological model if it gives good results in the description of a physical phenomenon. For instance, the usual Fourier diffusion equation is not invariant for time inversion, whereas the basic equations of classical mechanics are. Still, the Fourier equation gives very useful results when used in many applications. 

Finally, it is important to discuss some recent results by R. Hilfer \cite{hilfer00b,hilfer03b}. He has shown that not every continuous-time random walk with a long time tail is asymptotically equivalent to a diffusion equation with a fractional time derivative. In \cite{hilfer00b}, he considers different ways to define fractional derivatives in time. He shows that
only the Caputo type leads to mass-conserving fractional diffusion. In \cite{hilfer03b}, an example of a CTRW has been given whose waiting-time density has a power-law behaviour but whose diffusive limit is not the time-fractional diffusion equation. This latter counterexample seems to be in contrast with what has been said in Sec. IV above. However, the paradox disappears if one takes into account the proper scaling given by eqs. (\ref{scaledpdft}-\ref{scaledtransfs}). Indeed, the counterexample by Hilfer does not satisfy this scaling. More precisely, the non-relevance of this counterexample for the theory developed in Sec. IV can be stated as follows. The waiting-time density of the second model presented by Hilfer cannot be written in the form of eq. (\ref{scaledpdft}): $\psi_{r}(\tau) = \psi(\tau / r)/r$. Essentially, each of the two addends of Hilfer's density has a different scaling form. The scaling of eq. (\ref{scaledpdft}) has already been used by the present authors in ref. \cite{gorenflo02}. It was previously used by Feller in deriving the diffusion equation from the simple symmetric random walk \cite{feller68}, by Balakrishnan in 1985
\cite{balakrishnan85} and, in recent times, independently from the authors of this paper, by Uchaikin and Saenko \cite{uchaikin03}.

We are currently working to extend our approach to the coupled case, but this will be the subject of future papers.

\section*{ACKNOWLEDGEMENTS}

Rudolf Gorenflo and Francesco Mainardi appreciate the support of the EU ERASMUS-SOCRATES program for visits to Bologna and Berlin that, besides teaching, were also useful for this joint research project; moreover R.G. would like to acknowledge Rudolf Hilfer for the useful discussions after an invited lecture at ICA-1, Stuttgart University and for providing him a copy of his counterexample before publication. The authors wish to thank the referees for helping in making this paper clearer for non-mathematicians.

\end{document}